\documentstyle[aps,multicol,epsf]{revtex}
\input epsf
\draft
\title{An analysis of dynamical suppression of spontaneous emission}
\author{P. R. Berman 
\\ {\em Physics Department, University of Michigan, Ann Arbor, Michigan 48109-1120} }
\date{\today }
\begin{document}
\maketitle
\begin{abstract}
It has been shown recently [see, for example, S.-Y. Zhu and M. O. Scully,
Phys. Rev. Lett. {\bf 76}, 388 (1996)] that a dynamical suppression of
spontaneous emission can occur in a three-level system when an external
field drives transitions between a metastable state and {\em two} decaying
states. What is unusual in the decay scheme is that the decaying states are
coupled directly by the vacuum radiation field. It is shown that decay
dynamics required for total suppression of spontaneous emission necessarily
implies that the level scheme is isomorphic to a three-level lambda system,
in which the lower two levels are {\em both} metastable, and each is coupled
to the decaying state. As such, the total suppression of spontaneous
emission can be explained in terms of conventional dark states and coherent
population trapping.
\end{abstract}

\section{Introduction}

\begin{figure}[tb!]
\centering
\begin{minipage}{8.0cm}
\epsfxsize= 8 cm \epsfysize= 8 cm \epsfbox{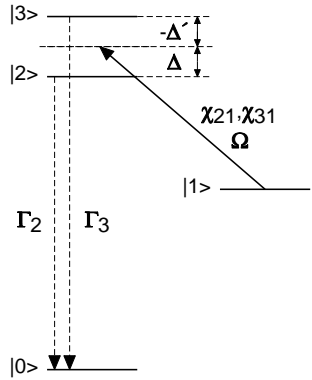}
\end{minipage}
\caption{Level scheme proposed by Zhu and Scully (see Ref. [4]) to
observe total suppression of spontaneous emission. The driving field having
frequency $\Omega $ couples state $\left| 1\right\rangle $ to both states $%
\left| 2\right\rangle $ and $\left| 3\right\rangle .$ with associated Rabi
frequencies $\chi _{21}$ and $\chi _{31}$, respectively. Spontaneous
emission is totally suppressed if $\Delta \chi _{31}^{2}+\Delta ^{\prime
}\chi _{21}^{2}=0$ where $\chi _{21}$ and $\chi _{31}$ are the Rabi
frequencies associated with the 1-2 and 1-3 transitions, respectively, and $%
\Delta \equiv \Omega -\omega _{21}$, $\Delta ^{\prime }\equiv \Omega -\omega
_{31}$.}
\label{fig1}
\end{figure}

Following the work of Fontana and Srivastava \cite{fontana}, Agarwal \cite
{ag}, Cardimona, Raymer, and Stroud \cite{card}, and Zhu and Scully \cite
{zhou}, a number of articles have appeared containing proposals for
suppressing spontaneous emission \cite
{gryn,sriv,nard,swain,agar,huang,lee,vlad}. In contrast to the suppression
of spontaneous emission that one can achieve by placing an atom in a cavity
whose radiation modes differ from those of free space, it is suggested in
these articles that spontaneous emission in free space can be suppressed by
applying an external radiation field to an atom having a specified level
scheme. This is a rather remarkable result, since one might imagine that,
owing to the very short correlation time of the vacuum field, such
modification of spontaneous emission rates would be strictly forbidden. A
prototypical level scheme that leads to suppression of spontaneous emission
is that of Zhu and Scully \cite{zhou} (see Fig. \ref{fig1}). Two excited states $%
\left| 2\right\rangle $ and $\left| 3\right\rangle $ are separated in
frequency by $\omega _{32}$. These states decay to the ground state $\left|
0\right\rangle $ with rates $\Gamma _{2}$ and $\Gamma _{3}$, respectively.
What makes the decay scheme somewhat unusual is that states $\left|
2\right\rangle $ and $\left| 3\right\rangle $ are {\em coupled directly by
the vacuum field}. An external radiation field couples an auxiliary, {\em %
metastable} state $\left| 1\right\rangle $ to both states $\left|
2\right\rangle $ and $\left| 3\right\rangle .$ For certain values of the
field strength and atom-field detunings, it is found that one can have a
nonvanishing, significant, steady state probability for the atom to be in
states $\left| 2\right\rangle $ or $\left| 3\right\rangle $. As such,
spontaneous emission from these levels is suppressed by the presence of the
driving field. Xia {\it et al. }\cite{xia} claim to have observed this
effect in an experiment on sodium dimers.

The suppression of spontaneous emission has been explained in terms of a
dressed state of the atom-field system that is decoupled from the vacuum
radiation field \cite{card,zhou,swain,huang,lee,xia}. How is this decoupling
accomplished? Is there any underlying structure in the proposed level
schemes that can help one to understand this most surprising result? It is
the purpose of this article to address these questions. By considering a
model problem, I will show that states $\left| 2\right\rangle $ and $\left|
3\right\rangle $ can be viewed as superpositions of two states, {\em one of
which is metastable}. It is this metastable state that is necessary for the
total suppression of spontaneous emission. Moreover, the decay dynamics
required for total suppression of spontaneous emission implies that the
level scheme of Fig. \ref{fig1} is isomorphic to a three-level lambda system. The
lower two levels of the lambda system are {\em both} metastable, and each is
coupled to the decaying state. As such, the total suppression of spontaneous
emission can be explained in terms of conventional dark states and coherent
population trapping \cite{ari}. The experiment of Xia et al \cite{xia} will
also be discussed. While the level scheme they study is relevant to this
class of problems, the results they obtained cannot be classified as a
suppression of spontaneous emission.

\section{Equations of Motion}

In the absence of the driving field, the equations for the evolution of the
state amplitudes $a_{2}$ and $a_{3}$ given by Zhu and Scully \cite{zhou} are

\begin{mathletters}
\label{1}
\begin{eqnarray}
\dot{a}_{2} &=&i(\omega _{32}/2)a_{2}-\gamma _{2}a_{2}-\gamma _{3,2}a_{3}
\label{1a} \\
\dot{a}_{3} &=&-i(\omega _{32}/2)a_{3}-\gamma _{3}a_{3}-\gamma _{3,2}a_{2}
\label{1b}
\end{eqnarray}
where 
\end{mathletters}
\begin{equation}
\gamma _{2}=\Gamma _{2}/2;\qquad \gamma _{3}=\Gamma _{3}/2;\qquad \gamma
_{3,2}=\sqrt{\gamma _{2}\gamma _{3}}.  \label{2}
\end{equation}

The first question we must ask is whether or not these equations correctly
describe the interaction of an atom with the vacuum radiation field. The
answer to this question is not obvious. If we consider the energy levels
shown in Fig. \ref{fig1} to be those of an isolated atom in free space, we
immediately run into some problems. From the dipole selection rules, it is
easy to show that the vacuum coupling from state $\left| 2\right\rangle $ to
state $\left| 3\right\rangle $ must conserve orbital, spin-orbit, and total
angular momenta ${\bf L}$, ${\bf J}$, and ${\bf F}$, as well as the
z-component of total angular momentum. As a consequence, states $\left|
2\right\rangle $ and $\left| 3\right\rangle $ {\em must belong to different
electronic state manifolds}$.$ This, in turn, implies that $\omega _{32}$
corresponds to a frequency that is orders of magnitude larger than the decay
rates $\Gamma _{2}$ and $\Gamma _{3}$, respectively. The rapid oscillation
of state amplitudes $a_{2}$ and $a_{3}$ with frequency $\omega _{32}$ brings
into question the validity of the Weisskopf-Wigner approximation used for
the derivation of Eqs. (\ref{1}). There is, perhaps, a more subtle point
involved. Starting from state $\left| 2\right\rangle $, one can emit a
photon taking the atom to state $\left| 0\right\rangle ,$ reabsorb this
photon taking the atom to {\em virtual state} $\left| 3\right\rangle ,$
reemit a photon taking the atom to state $\left| 0\right\rangle $ and
reabsorb this photon returning the atom to state $\left| 2\right\rangle $.
This overall process constitutes an $\alpha ^{5}($Rydberg) contribution to
the Lamb shift of state $\left| 2\right\rangle $. Consequently, if the
atomic states are renormalized to include the Lamb shift, it is questionable
as to whether the vacuum coupling in between states $\left| 2\right\rangle $
and $\left| 3\right\rangle $ should be included in Eqs. (\ref{1}) \cite{cs}.

It thus appears unlikely that one can achieve the vacuum coupling indicated
in Eqs. (\ref{1}) if these states correspond to eigenstates of a free,
isolated atom, dressed by the vacuum field. On the other hand, it {\em is}
possible to achieve this vacuum coupling if states $\left| 2\right\rangle $
and $\left| 3\right\rangle $ correspond to eigenstates of an atom plus some
external field or, in some cases, to the states of a molecule \cite{cs}. The
most obvious atom candidate is a hydrogen atom in a static electric field 
\cite{rf}. States $\left| 2\right\rangle $ and $\left| 3\right\rangle $
could then be chosen as linear combinations of the 2S and 3P states of
hydrogen. The idea of using a hydrogen atom in a static electric field to
modify the spontaneous emission spectrum is not new. Zhu and Scully mention
it in their 1996 article \cite{zhou}, and Fontana and Srivastava gave a
detailed analysis of the decay in their 1973 article \cite{fontana}.
Alternatively, one could use a level scheme similar to that used by Xia {\it %
et al. }\cite{xia}, in which states $\left| 2\right\rangle $ and $\left|
3\right\rangle $ are superposition of singlet and triplet states in a
molecule. I will return to the experiment of Xia {\it et al. }in Sec. III.

\begin{figure}[tb!]
\centering
\begin{minipage}{8.0cm}
\epsfxsize= 8 cm \epsfysize= 8 cm \epsfbox{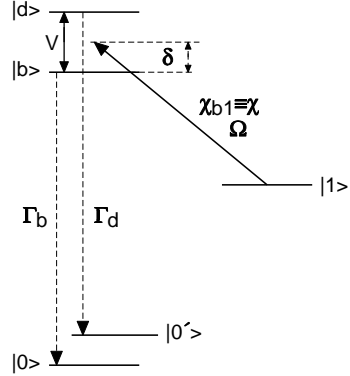}
\end{minipage}
\caption{A level scheme equivalent to that in Fig. 1 under conditions of
total suppression of spontaneous emission. A static field $V$ couples states 
$\left| b\right\rangle $ and $\left| d\right\rangle $ and a driving field
having frequency $\Omega $ couples state $\left| 1\right\rangle $ to state $%
\left| b\right\rangle $ only, with associated Rabi frequency $\chi $.
Spontaneous emission is totally suppressed if the detuning $\delta \equiv
\Omega -\omega _{b1}=\omega _{db}$ and $\gamma _{d}=0$.}
\label{fig2}
\end{figure}

In order to gain additional insight into this problem, I consider the level
scheme shown in Fig. \ref{fig2}. States $\left| 0\right\rangle $, $\left| 0^{\prime
}\right\rangle ,$ $\left| b\right\rangle $, $\left| d\right\rangle $, and $%
\left| 1\right\rangle $ are eigenstates of an unperturbed Hamiltonian.
States $\left| b\right\rangle $ and $\left| d\right\rangle $ have opposite
parity and are coupled by a constant potential $\hbar V$ \cite{rf}. An
external radiation field, having frequency $\Omega ,$ couples state $\left|
1\right\rangle $, which is assumed to be metastable, to state $\left|
b\right\rangle $ only. States $\left| b\right\rangle $ and $\left|
d\right\rangle $ decay to states $\left| 0\right\rangle $ and $\left|
0^{\prime }\right\rangle $ with rates $\Gamma _{b}$ and $\Gamma _{d}$,
respectively. (Note that this level scheme could correspond to hydrogen with
state $\left| b\right\rangle $ corresponding to $\left| n=2,L=1,m_{\ell
}=0\right\rangle $, state $\left| d\right\rangle $ to $\left|
n=2,L=0,m_{\ell }=0\right\rangle $, and states $\left| 0\right\rangle $, $%
\left| 0^{\prime }\right\rangle $ and $\left| 1\right\rangle $ to $\left|
n=1,L=0,m_{\ell }=0\right\rangle $ \cite{decay}. In this case, $\Gamma
_{d}\approx 0$ for the $2S$ state$.$) The goal of this calculation is to
show that the level scheme of Fig. \ref{fig2}, a level scheme exhibiting conventional
decay dynamics, can be mapped into that of Fig. \ref{fig1}, a level scheme exhibiting
somewhat unconventional decay dynamics. Thus, the suppression of spontaneous
emission can equally well be analyzed using the level schemes of Fig. \ref{fig1} or
\ref{fig2}. It will be seen that the suppression of spontaneous emission can be
explained in terms of conventional dark states when the level scheme of Fig.
\ref{fig2} is used. The mapping between the two level schemes is achieved by
identifying states $\left| 2\right\rangle $ and $\left| 3\right\rangle $,
appearing in Fig. \ref{fig1}, as eigenstates of the unperturbed Hamiltonian
associated with Fig. \ref{fig2}, plus the potential $\hbar V.$

In the absence of decay, the effective Hamiltonian for the level scheme of
Fig. \ref{fig2}, in the rotating wave approximation, in an interaction
representation, and with the energy of level $b$ taken equal to zero, can be
written as \cite{sal} 
\begin{equation}
{\bf H}_{0}=\hbar \left( 
\begin{array}{lll}
\delta  & \chi  & 0 \\ 
\chi  & 0 & V \\ 
0 & V & \delta ^{\prime }
\end{array}
\right) ,  \label{3}
\end{equation}
where the order of the states is ($\left| 1\right\rangle $,$\left|
b\right\rangle $,$\left| d\right\rangle )$, $\chi $ is a Rabi frequency
(taken to be real) and \cite{rf} 
\begin{equation}
\delta =\Omega -\omega _{b1};\qquad \delta ^{\prime }\equiv \omega _{db}.
\label{4}
\end{equation}
States $\left| 0\right\rangle $ and $\left| 0^{\prime }\right\rangle $ have
not been included in (\ref{3}) since they are not needed for the present
discussion of the decay dynamics of the excited states.

The Hamiltonian (\ref{3}) can be diagonalized without much difficulty;
however, the desired comparison between the level schemes of Figs. 1 and 2
is achieved by diagonalizing the $b,d$ subspace only. The new eigenstates
are given by 
\begin{mathletters}
\label{5}
\begin{eqnarray}
\left| 2\right\rangle &=&c\left| b\right\rangle -s\left| d\right\rangle
\label{5a} \\
\left| 3\right\rangle &=&c\left| d\right\rangle +s\left| b\right\rangle
\label{5b}
\end{eqnarray}
where 
\begin{equation}
c\equiv \cos \theta =\sqrt{\frac{1}{2}\left( 1+\frac{\delta ^{\prime }}{R_{A}%
}\right) };\qquad s\equiv \sin \theta ;\qquad \tan 2\theta =2V/\delta
^{\prime };\qquad R_{A}=\sqrt{\delta ^{\prime 2}+4V^{2}}.  \label{5p}
\end{equation}

In terms of these eigenstates, the transformed Hamiltonian takes the form 
\end{mathletters}
\begin{equation}
{\bf H}_{0}^{\prime }=\hbar \left( 
\begin{array}{lll}
\delta -\frac{\delta ^{\prime }}{2} & c\chi & s\chi \\ 
c\chi & \frac{-R_{A}}{2} & 0 \\ 
s\chi & 0 & \frac{R_{A}}{2}
\end{array}
\right) ,  \label{7}
\end{equation}
where the order of the states is ($\left| 1\right\rangle $,$\left|
2\right\rangle $,$\left| 3\right\rangle )$ and a constant energy $\hbar
\delta ^{\prime }/2$ has been subtracted from the energy of each state. The
equations of motion for the state amplitudes are 
\begin{mathletters}
\label{9}
\begin{eqnarray}
\dot{a}_{1} &=&-i\left( \delta -\frac{\delta ^{\prime }}{2}\right)
a_{1}-ic\chi a_{2}-is\chi a_{3}  \label{9a} \\
\dot{a}_{2} &=&i(R_{A}/2)a_{2}-ic\chi a_{1}  \label{9b} \\
\dot{a}_{3} &=&-i(R_{A}/2)a_{3}-is\chi a_{1}  \label{9c}
\end{eqnarray}
It is now a simple matter to include decay into these equations. Since
spontaneous decay is governed by $\dot{a}_{b}=-\gamma _{b}a_{b}$; $\dot{a}%
_{d}=-\gamma _{d}a_{d},$ where 
\end{mathletters}
\begin{equation}
\gamma _{b}=\Gamma _{b}/2;\qquad \gamma _{d}=\Gamma _{d}/2,  \label{10}
\end{equation}
and since $a_{2}=ca_{b}-sa_{d}$; $a_{3}=ca_{d}+sa_{b}$, it follows that Eqs.
(\ref{9}), including decay, can be written as 
\begin{mathletters}
\label{11}
\begin{eqnarray}
\dot{a}_{1} &=&-i\left( \delta -\frac{\delta ^{\prime }}{2}\right)
a_{1}-ic\chi a_{2}-is\chi a_{3}  \label{11a} \\
\dot{a}_{2} &=&-\gamma _{2}a_{2}-\gamma _{3,2}a_{3}+i(\omega
_{32}/2)a_{2}-ic\chi a_{1}  \label{11b} \\
\dot{a}_{3} &=&-\gamma _{3}a_{3}-\gamma _{2,3}a_{2}-i(\omega
_{32}/2)a_{3}-is\chi a_{1}  \label{11c}
\end{eqnarray}
where 
\end{mathletters}
\begin{mathletters}
\label{12}
\begin{eqnarray}
\gamma _{2} &=&c^{2}\gamma _{b}+s^{2}\gamma _{d}  \label{12a} \\
\gamma _{3} &=&c^{2}\gamma _{d}+s^{2}\gamma _{b}  \label{12b} \\
\gamma _{3,2} &=&\gamma _{2,3}=sc(\gamma _{b}-\gamma _{d})  \label{12c} \\
\omega _{32} &=&R_{A}.  \label{12d}
\end{eqnarray}

This form of the equations is {\em almost }identical to that used in
theories of suppression of spontaneous decay [compare with Eq. (\ref{1})]
based on the level scheme of Fig. \ref{fig1}. For the equations to be{\em \
identical, }and for the level schemes of Figs. 1 and 2 to be isomorphic, one
must require that 
\end{mathletters}
\begin{equation}
\gamma _{3,2}=\sqrt{\gamma _{2}\gamma _{3}}.  \label{13}
\end{equation}
It follows from Eqs. (\ref{12}) that the only way this equation can be
satisfied is to have $\gamma _{d}=0$. In other words, the form of the vacuum
coupling given in Eqs. (\ref{1}) for the level scheme of Fig. \ref{fig1} in theories
of total suppression of spontaneous {\em necessarily implies that state }$%
\left| d\right\rangle $ {\em of the equivalent level scheme} {\em of Fig. \ref{fig2}} 
{\em must be metastable}.

Since both states $\left| d\right\rangle $ and $\left| 1\right\rangle $ are
metastable and do not undergo spontaneous emission in the isolated atom, it
is reasonable to ask whether or not the level scheme of Fig. \ref{fig1} legitimately
qualifies to be labeled as one in which spontaneous emission has been
suppressed. In order to determine if the driving field suppresses
spontaneous emission, one must first establish that spontaneous emission of
states $\left| 2\right\rangle $ and $\left| 3\right\rangle $ always occurs
in the {\em absence} of the driving field. Setting $\chi =0$ in Eqs. (\ref
{11}), one finds that the only steady state solution is $a_{2}=a_{3}=0.$ Any
initial state population in states $\left| 2\right\rangle $ and $\left|
3\right\rangle $ decays if the driving field is absent. This is easily
understood in terms of the original $\left| b\right\rangle $, $\left|
d\right\rangle $ basis; although state $\left| d\right\rangle $ is
metastable, it is coupled to the decaying state $\left| b\right\rangle $ by
the potential $\hbar V.$ No matter how weak the coupling strength $V$, any
initial population in state $\left| d\right\rangle $ eventually leaks out
via state $\left| b\right\rangle .$

Does the presence of the driving field suppress this spontaneous emission?
The answer to this question is affirmative if the initial state is an
arbitrary superposition of states $\left| 2\right\rangle $ and $\left|
3\right\rangle $ and their remains population trapped in states $\left|
2\right\rangle $ and $\left| 3\right\rangle $ as the time approaches
infinity. An initial condition in which the atom is in state $\left|
1\right\rangle ,$ corresponding to the initial condition in the experiment
of Xia {\it et al. }\cite{xia}, cannot be used directly to establish total
suppression of spontaneous emission, but can provide indirect evidence for
this effect, as discussed in Sec. III below.

Having established that state $\left| d\right\rangle $ must be metastable to
satisfy the requirements for spontaneous emission suppression, it is now an
easy matter to understand the total suppression of spontaneous emission by
returning to the original Hamiltonian (\ref{3}). An inspection of this
Hamiltonian reveals that it is identical to a Hamiltonian, written in a
field interaction representation \cite{sal}, that characterizes a
three-level atom in a lambda scheme driven by two fields. The field having
Rabi frequency $\chi $ and detuning $\delta =\Omega -\omega _{b1}$ drives
the $1$-$b$ transition and the field having Rabi frequency $V$ and detuning $%
\delta ^{\prime }\equiv \omega _{bd}$ drives the $b$-$d$ transition \cite{rf}%
. Total suppression of spontaneous emission occurs if one can find an
eigenstate consisting of a superposition of state amplitudes $a_{1}$ and $%
a_{d}$ which is decoupled from state amplitude $a_{b}$. This {\em dark state}
\cite{ari} does not decay since it is a superposition of nondecaying states.
In other words we seek values of $\alpha $ and $\beta $ for which the
superposition of state amplitudes of the form 
\begin{equation}
a_{I}=\alpha a_{1}+\beta a_{d}  \label{15}
\end{equation}
satisfies the equation of motion 
\begin{equation}
\dot{a}_{I}=-i\omega _{I}a_{I}.  \label{16}
\end{equation}
From Schr\"{o}dinger's Equation with the Hamiltonian (\ref{3}), it follow
that 
\begin{equation}
\dot{a}_{I}=\alpha \dot{a}_{1}+\beta \dot{a}_{d}=-i(\alpha \chi +\beta
V)a_{b}-i(\alpha \delta a_{1}+\beta \delta ^{\prime }a_{d}).  \label{17}
\end{equation}
Equation (\ref{16}) can be satisfied only if 
\begin{equation}
\alpha \chi +\beta V=0;  \label{17b}
\end{equation}
\begin{equation}
\delta =\delta ^{\prime }\equiv \omega _{db},  \label{18}
\end{equation}
which implies that $\omega _{I}=\delta $. The driving field must be tuned to
the frequency that would correspond to a ''hole'' in the emission spectrum
from the 2-3 state manifold \cite{fontana}. Thus if $\delta =\omega _{bd}$
there always exists a dark state amplitude of the system 
\begin{equation}
a_{I}=\frac{Va_{1}-\chi a_{d}}{R_{B}},  \label{19q}
\end{equation}
where
\begin{equation}
R_{B}=\sqrt{V^{2}+\chi ^{2}},  \label{19s}
\end{equation}
which does not decay. The other eigenstate amplitudes 
\begin{mathletters}
\label{14}
\begin{eqnarray}
a_{II} &=&\frac{\chi a_{1}-(R_{D}+\delta /2)a_{b}+Va_{d}}{\sqrt{%
(R_{D}+\delta /2)^{2}+R_{B}^{2}}};  \label{20a} \\
a_{III} &=&\frac{\chi a_{1}+(R_{D}-\delta /2)a_{b}+Va_{d}}{\sqrt{%
(R_{D}+\delta /2)^{2}+R_{B}^{2}}},  \label{20c}
\end{eqnarray}
where
\end{mathletters}
\begin{equation}
R_{D}=\sqrt{R_{B}^{2}+(\delta /2)^{2}},  \label{20s}
\end{equation}
contain an admixture of state amplitude $a_{b}$ and decay as $t\sim \infty $%
. As a consequence, any initial condition for which $a_{I}(0)\neq 0$ has a
metastable component that does not decay as the time approaches infinity.

It remains only to establish that an initial condition of the form $\left|
\psi (0)\right\rangle =a_{2}(0)\left| 2\right\rangle +a_{3}(0)\left|
3\right\rangle $ leads to a final state which has some population trapped in
states $\left| 2\right\rangle $ and $\left| 3\right\rangle $ (or,
equivalently, in state $\left| d\right\rangle ).$ As $t\sim \infty $, the
solution for the total dressed state amplitudes $a_{I},a_{II},a_{III}$ is $%
a_{I}(t)\sim a_{I}(0)e^{-i\delta t},$ $a_{II}(t)\sim 0,$ $a_{III}(t)\sim 0$,
which, when reexpressed in terms of the bare state initial conditions [with $%
a_{1}(0)=0]$ is $a_{I}(t)\sim -(\chi /R_{B})a_{d}(0)e^{-i\delta t}$. The
final state populations are 
\begin{mathletters}
\label{21}
\begin{eqnarray}
\left| a_{1}(\infty )\right| ^{2} &=&\left( \frac{\chi V}{V^{2}+\chi ^{2}}%
\right) ^{2}\left| a_{d}(0)\right| ^{2}  \label{21a} \\
\left| a_{b}(\infty )\right| ^{2} &=&0  \label{21b} \\
\left| a_{d}(\infty )\right| ^{2} &=&\left( \frac{\chi ^{2}}{V^{2}+\chi ^{2}}%
\right) ^{2}\left| a_{d}(0)\right| ^{2}  \label{21c} \\
\left| a_{2}(\infty )\right| ^{2} &=&s^{2}\left| a_{d}(\infty )\right| ^{2}=%
\frac{1}{2}\left( 1-\frac{\omega _{db}}{\sqrt{\omega _{db}^{2}+4V^{2}}}%
\right) \left( \frac{\chi ^{2}}{V^{2}+\chi ^{2}}\right) ^{2}\left|
a_{d}(0)\right| ^{2}  \label{21d} \\
\left| a_{3}(\infty )\right| ^{2} &=&c^{2}\left| a_{d}(\infty )\right| ^{2}=%
\frac{1}{2}\left( 1+\frac{\omega _{db}}{\sqrt{\omega _{db}^{2}+4V^{2}}}%
\right) \left( \frac{\chi ^{2}}{V^{2}+\chi ^{2}}\right) ^{2}\left|
a_{d}(0)\right| ^{2}  \label{21e}
\end{eqnarray}
\end {mathletters}
where Eqs. (\ref{4}), (\ref{5p}) and (\ref{19s}) were used. Thus, we see
that population is always trapped in states $\left| 2\right\rangle $ and $%
\left| 3\right\rangle .$

Equations (\ref{21}) for the probabilities $\left| a_{1}(\infty )\right| ^{2}
$, $\left| a_{2}(\infty )\right| ^{2}$, and $\left| a_{3}(\infty )\right|
^{2}$ can be written in terms of the couplings and detunings in the ($\left|
1\right\rangle $,$\left| 2\right\rangle $,$\left| 3\right\rangle )$ basis.
Referring to Eq. (\ref{7}) and using Eqs. (\ref{4}), (\ref{5p}), (\ref{12d}%
), (\ref{18}), (\ref{19s}), one finds the appropriate relationships, $\omega
_{db}=(\Delta +\Delta ^{\prime });$ $V=\sqrt{-\Delta \Delta ^{\prime }};$ $%
\chi ^{2}=\chi _{21}^{2}+\chi _{31}^{2}$, and $a_{d}=ca_{3}-sa_{2};$ $c=%
\sqrt{\Delta /\omega _{32}};$ $s=\sqrt{-\Delta ^{\prime }/\omega _{32}}$,
subject to the constraint, $\Delta \chi _{31}^{2}+\Delta ^{\prime }\chi
_{21}^{2}=0$. Under conditions of total suppression of spontaneous emission,
the field is tuned to the energy of the metastable level $d$; that is, $%
\Delta =\omega _{d2}>0$ and $\Delta ^{\prime }=\omega _{d3}<0$. 

As an aside, I might note that the results can be reinterpreted as a
suppression of absorption \cite{card,absorp} if one starts with all
population initially in state $\left| 1\right\rangle $, for which $%
a_{I}(t)\sim -(V/R_{B})a_{1}(0)e^{-i\delta t}$ as $t\sim \infty $. In the
absence of the coupling potential $\hbar V$, the steady state population $%
\left| a_{1}(\infty )\right| ^{2}$ would vanish, but it does not vanish in
the presence of this coupling. In this sense, it is closely related to
electromagnetically induced transparency \cite{harris}.

\section{Discussion}

It has been shown that the origin of the suppression of spontaneous emission
proposed by Zhu and Scully \cite{zhou} and others \cite
{nard,swain,huang,lee,vlad} can be traced to a metastable state that is
''hidden'' in their calculations. Once this hidden state is revealed, the
suppression of spontaneous emission can be understood in terms of a
conventional dark state and coherent population trapping \cite{ari} that can
arise when an atom having a three-level, lambda scheme is driven by two
fields. The dark state in this instance is a superposition of two metastable
states so is, itself, metastable. The dynamical suppression of spontaneous
emission is a real effect. If the external driving field $\chi $ were not
present, the two state manifold consisting of states $\left| 2\right\rangle $
and $\left| 3\right\rangle $ would always decay. In some sense, the driving
field allows one to access the metastable level $\left| d\right\rangle $
contained in both states $\left| 2\right\rangle $ and $\left| 3\right\rangle 
$. This type of dynamical suppression could be used, for example, to reduce
spontaneous emission in the $2S$-$2P$ manifold of hydrogen resulting from
stray fields that couple the $S$ state to the $P$ state. It would be
necessary to drive the $2S$-$2P$ transition using an {\it rf }field and the $%
1S$-$2P$ with a uv field having frequency $\Omega =\omega _{2S,1S}-\Omega
_{rf}$ \cite{rf}.

The use of the equivalent ($\left| 1\right\rangle ,\left| b\right\rangle
,\left| d\right\rangle )$ basis rather than the ($\left| 1\right\rangle
,\left| 2\right\rangle ,\left| 3\right\rangle )$ basis greatly simplifies
the interpretation of the results. From the analysis of Sec. II, it is clear
that the final state probabilities depend only on the initial state
amplitude $a_{I}(0)=\left[ Va_{1}(0)-\chi a_{d}(0)\right] /R_{B}$ and not on
the decay rate if state $\left| b\right\rangle $ decays to state $\left|
0\right\rangle $ only. If state $\left| b\right\rangle $ decays to state $%
\left| 1\right\rangle $ as well as to state $\left| 0\right\rangle $, or if
states $\left| 1\right\rangle $ and $\left| 0\right\rangle $ actually
correspond to the same state$,$ the final state probabilities are modified,
but the steady state still corresponds to a dark state for which there is
total suppression of absorption. On the other hand, if state $\left|
1\right\rangle $ is not metastable, there cannot be total suppression of
spontaneous emission since the dark state amplitude $a_{I}(t)=\left[
Va_{1}(t)-\chi a_{d}(t)\right] /R_{B}$ decays to zero as $t\sim \infty $.

\begin{figure}[tb!]
\centering
\begin{minipage}{8.0cm}
\epsfxsize= 8 cm \epsfysize= 8 cm \epsfbox{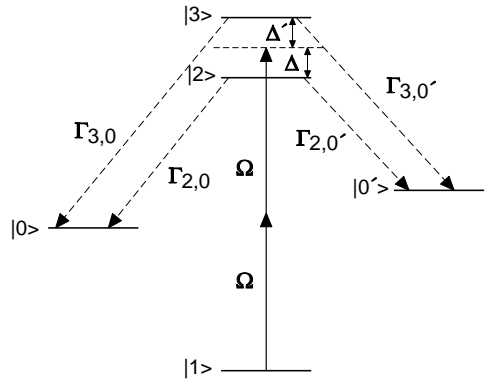}
\end{minipage}
\caption{The level scheme used in the experiment of Xia {\it et al.} Ref. [13].}
\label{fig3}
\end{figure}

Finally I should like to discuss the experiment of Xia {\it et al. }\cite
{xia}{\it , }who used the level scheme shown in Fig. \ref{fig3}, corresponding to
molecular states in the sodium dimer. States $\left| 2\right\rangle $ and $%
\left| 3\right\rangle $ are superpositions of singlet and triplet states
which are mixed by a spin-orbit interaction. In the spirit of this
calculation, one can associate the singlet and triplet states with states $%
\left| b\right\rangle $ and $\left| d\right\rangle ,$ respectively, in the
Hamiltonian (\ref{3}), and the spin-orbit mixing with the potential $\hbar V$%
. Of course, it is not possible to ''turn off'' the mixing potential in this
case. The singlet component of states $\left| 2\right\rangle $ and $\left|
3\right\rangle $ decays to state $\left| 0\right\rangle $ and the triplet
component of states $\left| 2\right\rangle $ and $\left| 3\right\rangle $
decays to state $\left| 0^{\prime }\right\rangle $, while the singlet
component of states $\left| 2\right\rangle $ and $\left| 3\right\rangle $ is
driven by a two-photon transition from the ground state. Since {\em both}
the singlet and triplet components decay, the conditions for suppression of
spontaneous emission are not met (recall that it was necessary that state $%
\left| d\right\rangle ,$ which corresponds to the triplet state, be
metastable). 
In their experiment, Xia {\it et al. }are not measuring spontaneous{\it \ }%
emission, as it is normally defined. Instead, they are measuring {\em %
scattering} via the {\em three-photon} process in which two photons are
absorbed from the driving field and a vacuum photon is emitted taking the
atom to state $\left| 0\right\rangle $ (singlet channel) or $\left|
0^{\prime }\right\rangle $ (triplet channel). They found that, for a tuning
of the incident field midway between levels $\left| 2\right\rangle $ and $%
\left| 3\right\rangle $, 2$\Omega =\left( \omega _{31}+\omega _{31}\right) /2
$, the scattering in the singlet channel was suppressed and that in the
triplet channel was enhanced. This constitutes strong evidence that states $%
\left| 2\right\rangle $ and $\left| 3\right\rangle $ are coupled directly by
the vacuum field, and that the singlet and triplet states are degenerate in
the absence of the spin-orbit coupling (${\it i.e.,}$ $\omega _{db}=0$) \cite
{scat}. Although this experiment is important insofar as it provides an
example of a system in which vacuum coupling of two, distinct excited states
occurs, it does not demonstrate suppression of spontaneous emission. There
will be no steady-state population in states $\left| 2\right\rangle $ and $%
\left| 3\right\rangle $. On the other hand, XIA {\it et al.} have shown that
scattering in a specific channel can be suppressed.

As was noted above, total suppression of absorption occurs under the same
conditions as total suppression of spontaneous emission, so that the
existence of one implies the other. Consequently, if one can demonstrate
total suppression of absorption, the system will also exhibit total
suppression of spontaneous emission. To establish total suppression of
absorption, one can either (i) prove that there is no absorption of the
driving field or (ii) show that there is no scattered radiation for {\em all}
polarizations and directions of the scattered field. The existence of
scattered radiation in the triplet channel in the experiment of Xia {\it et
al. }necessarily implies that there is {\em not} total suppression of
absorption. On the other hand, the absorption rate from the ground state is
decreased by a factor $2\gamma _{d}$/($\gamma _{d}+\gamma _{b}$) relative to
that which would have occurred if states $\left| 2\right\rangle $ and $%
\left| 3\right\rangle $ were not coupled by the vacuum field. Consequently,
one can say that the spontaneous emission rate or the absorption rate is
decreased in this system if $\gamma _{d}($triplet)$\ll \gamma _{b}($%
singlet). The data seems to indicate that $\gamma _{d}$ and $\gamma _{b}$
are comparable.

\section{Acknowledgments}

I am pleased to acknowledge helpful discussions with J. L. Cohen, B.
Dubetsky, P. Milonni, and G. W. Ford. This work is supported by the National
Science Foundation under Grant No. PHY-9414020 and the U. S. Office of Army
Research under Grant No. DAAG55-97-0113.


\begin{references}
\bibitem{decay}  Suppression of spontaneous emission still occurs if states $%
\left| 0\right\rangle $ and $\left| 1\right\rangle $ correspond to the same
physical state (see Sec. III).
\end{references}
\end{document}